\documentclass[epj]{webofc}
\usepackage[varg]{txfonts}   
\usepackage{graphicx}
\usepackage{dcolumn}
\usepackage{bm}
\usepackage{amssymb}
\usepackage{amsfonts}
\usepackage{amsmath}
\usepackage{latexsym}
\usepackage{dsfont}
\usepackage{multirow}
\usepackage{color}
\usepackage[mathscr]{eucal}

\newcommand{\beq}{\begin{equation}}
\newcommand{\eeq}{\end{equation}}
\newcommand{\beqy}{\begin{eqnarray}}
\newcommand{\eeqy}{\end{eqnarray}}
\newcommand{\beqyn}{\begin{eqnarray*}}
\newcommand{\eeqyn}{\end{eqnarray*}}

\newcommand{\bs}{\begin{slide}}
\newcommand{\es}{\end{slide}}
\newcommand{\bc}{\begin{center}}
\newcommand{\ec}{\end{center}}
\newcommand{\bmin}{\begin{minipage}}
\newcommand{\emin}{\end{minipage}}

\newcommand{\bi}{\begin{itemize}}
\newcommand{\ei}{\end{itemize}}

\newcommand{\bea}{\begin{eqnarray}}
\newcommand{\eea}{\end{eqnarray}}
\newcommand{\be}{\begin{equation}}
\newcommand{\ee}{\end{equation}}

\newcommand{\ud}{\mathrm{d}}

\newcommand{\uTr}{\mathrm{Tr}}

\newcommand{\uD}{\mathcal{D}}

\newcommand{\barpsi}{\overline{\psi}}

\newcommand{\pure}{\text{pure}}
\newcommand{\phys}{\text{phys}}

\newcommand{\LRpartial}{\overset{\leftrightarrow}{\partial}\!\!\!\!\phantom{\partial}}
\newcommand{\LRD}{\overset{\leftrightarrow}{D}\!\!\!\!\!\phantom{D}}
\woctitle{POETIC 2015}
\begin{document}
\title{The gauge-invariant canonical energy-momentum tensor}
%
%

\author{C\'edric Lorc\'e\inst{1}\fnsep\thanks{\email{cedric.lorce@polytechnique.edu}}
}

\institute{Centre de Physique Th\'eorique, Ecole polytechnique, CNRS, Universit\'e Paris-Saclay, F-91128 Palaiseau, France 
 }

\abstract{%
The canonical energy-momentum tensor is often considered as a purely academic object because of its gauge dependence. However, it has recently been realized that canonical quantities can in fact be defined in a gauge-invariant way provided that strict locality is abandoned, the non-local aspect being dictacted in high-energy physics by the factorization theorems. Using the general techniques for the parametrization of non-local parton correlators, we provide for the first time a complete parametrization of the energy-momentum tensor (generalizing the purely local parametrizations of Ji and Bakker-Leader-Trueman used for the kinetic energy-momentum tensor) and identify explicitly the parts accessible from measurable two-parton distribution functions (TMDs and GPDs). As by-products, we confirm the absence of model-independent relations between TMDs and parton orbital angular momentum, recover in a much simpler way the Burkardt sum rule and derive three similar new sum rules expressing the conservation of transverse momentum.
}
\maketitle
\section{Introduction}

Following Noether's procedure, one obtains a canonical energy-momentum (EMT) $T^{\mu\nu}$ which is usually neither symmetric\footnote{We stress that the symmetry requirement essentially comes from General Relativity where torsion is assumed to vanish. In more general theories of gravitation like \emph{e.g.} Einstein-Cartan theory and metric-affine gauge theory, torsion is allowed to be nonzero leading to asymmetric EMTs and a natural coupling between gravitation and spin. These effects are however expected to be extremely small and to show up only under extreme conditions, see \emph{e.g.}~\cite{Obukhov:2014fta} and references therein.} nor gauge invariant. These properties are often considered as pathological and can be cured by adding to the EMT a superpotential term of the form $\partial_\alpha f^{[\alpha\mu]\nu}$~\cite{Belinfante:1939,Belinfante:1940,Rosenfeld:1940}, where the square brackets stand for antisymmetrization. This amounts to a redefinition of the local density of energy and momentum~\cite{Hehl:1976vr,Liu:2015xha} while leaving the total linear and angular momenta unchanged. 

Superpotential terms have also been used to decompose the angular momentum into spin and orbital contributions~\cite{Jaffe:1989jz,Ji:1996ek}. According to standard textbooks like \emph{e.g.}~\cite{Cohen,Simmons}, one cannot perform this decomposition for the gauge field in a gauge-invariant way. Nonetheless, it appears that the photon spin and orbital angular momentum (OAM) are routinely measured, see \emph{e.g.}~\cite{Bliokh:2014ara} and references therein. Similarly, a gauge-invariant quantity called $\Delta G$, interpreted in the light-front gauge as the gluon spin~\cite{Jaffe:1989jz}, has been measured in polarized deep inelastic and proton-proton scatterings, see~\cite{deFlorian:2014yva} for a recent analysis. Since physical observables are gauge invariant, Chen \emph{et al.}~\cite{Chen:2008ag} claimed that the textbooks were wrong and proposed a formal gauge-invariant decomposition of the photon and gluon angular momentum. Their work received strong criticisms and triggered numerous theoretical papers on the subject, summarized in recent reviews~\cite{Leader:2013jra,Wakamatsu:2014zza}. It is now understood that textbooks implicitly referred to local quantities only, whereas the formal construction of Chen \emph{et al.} and the above-mentioned quantities extracted from experimental data are intrinsically non-local~\cite{Hatta:2011zs,Lorce:2012rr,Lorce:2012ce,Lorce:2013bja}. 

Typical examples of measurable non-local quantities in Quantum ChromoDynamics (QCD) are parton distribution functions (PDFs) whose gauge invariance is ensured by a Wilson line along a path determined by the factorization theorems~\cite{Collins:2011zzd}. Generalized Transverse-Momentum dependent Distributions (GTMDs)~\cite{Meissner:2009ww} are natural generalizations of the PDFs and provide a natural way to access the canonical OAM~\cite{Lorce:2011kd,Lorce:2011ni,Hatta:2011ku} and other angular momentum correlations~\cite{Lorce:2015sqe}. Unfortunately, it is not known so far how to access them directly in experiments. They can however be accessed indirectly using \emph{e.g.} realistic models~\cite{Lorce:2011kd,Kanazawa:2014nha,Mukherjee:2014nya,Mukherjee:2015aja,Liu:2015eqa}, or lattice QCD in the infinite-momentum limit~\cite{Musch:2010ka,Musch:2011er,Ji:2013dva,Ji:2014lra}.

We present here the first complete parametrization of the EMT obtained in~\cite{Lorce:2015lna} which can be applied to virtually any form of the EMT discussed in the literature, provided that the non-locality appears only along the light-front (LF) direction $n$.

\section{A basis of gauge-invariant energy-momentum tensors}\label{sec2}

Most of the EMTs that appeared in the literature can be decomposed in a basis of five gauge-invariant tensors
\begin{equation}
\begin{aligned}
T^{\mu\nu}_1&=\barpsi\gamma^\mu \tfrac{i}{2}\LRD^\nu\psi,&
T^{\mu\nu}_2&=-2\uTr\!\left[G^{\mu\alpha}G^\nu_{\phantom{\nu}\alpha}\right]+g^{\mu\nu}\,\tfrac{1}{2}\uTr\!\left[G^{\alpha\beta}G_{\alpha\beta}\right]\!,\\
T^{\mu\nu}_3&=-\barpsi\gamma^\mu gA^\nu_\phys\psi,&
T^{\mu\nu}_4&=\tfrac{1}{4}\,\epsilon^{\mu\nu\alpha\beta}\partial_\alpha\!\left[\barpsi\gamma_\beta\gamma_5\psi\right]\!,\\
T^{\mu\nu}_5&=-2\partial_\alpha\uTr\!\left[G^{\mu\alpha}A^\nu_\phys\right]\!,
\end{aligned}
\end{equation}
where $\tfrac{i}{2}\LRD^\mu=\tfrac{i}{2}\LRpartial^\mu+gA^\mu$ is the hermitian covariant derivative with $\LRpartial^\mu=\overset{\rightarrow}{\partial}\!\!\!\!\phantom{\partial}^\mu-\overset{\leftarrow}{\partial}\!\!\!\!\phantom{\partial}^\mu$, and $\epsilon_{0123}=+1$. For example, the kinetic form of the quark and gluon EMTs are given by $T^{\mu\nu}_1$ and $T^{\mu\nu}_2$, respectively, and the corresponding canonical forms are given by $T^{\mu\nu}_1+T^{\mu\nu}_3$ and $T^{\mu\nu}_2-T^{\mu\nu}_3+T^{\mu\nu}_5$, respectively. The superpotentials $T^{\mu\nu}_4$ and $T^{\mu\nu}_5$ together with the QCD equations of motion
\begin{equation}\label{QCDEOM}
\begin{aligned}
\barpsi(r)\gamma^{[\mu} i\LRD^{\nu]}\psi&=-\epsilon^{\mu\nu\alpha\beta}\partial_\alpha\!\left[\barpsi\gamma_\beta\gamma_5\psi\right]\!,\\
2\!\left[\uD_\alpha G^{\alpha\beta}\right]^c_{\phantom{c}c'}&=-g\,\barpsi_{c'}\gamma^\beta\psi^c,
\end{aligned}
\end{equation}
where $c,c'$ are color indices in the fundamental representation and $\mathcal D_\mu=\partial_\mu-ig[A_\mu,\quad]$ is the adjoint covariant derivative, can be used to relate the various EMTs~\cite{Leader:2013jra,Lorce:2015lna}. Because of the first identity in Eq.~\eqref{QCDEOM}, we have $T^{\mu\nu}_4=-\tfrac{1}{2}\,T^{[\mu\nu]}_1$ which can then be discarded in the following discussions. 

In order to obtain gauge-invariant canonical EMTs, one needs to decompose the gauge potential $A_\mu$ into pure-gauge and ``physical'' (or covariant) contributions
\begin{equation}
A^\pure_\mu\equiv\tfrac{i}{g}\,\mathcal W\partial_\mu\mathcal W^{-1},\qquad A^\phys_\mu\equiv A_\mu-A^\pure_\mu,
\end{equation}
where $\mathcal W$ is a non-integrable phase factor transforming as $\mathcal W\mapsto U\mathcal W$ under gauge transformations. In the gauge where $\mathcal W(r)=1$, the gauge-invariant canonical decomposition formally reduces to the Jaffe-Manohar decomposition~\cite{Jaffe:1989jz}, and can therefore be considered as a gauge-invariant extension of the latter~\cite{Ji:2012ba,Lorce:2012rr,Lorce:2013gxa,Leader:2013jra}.

\section{General parametrization}\label{sec3}

Since our aim is to relate the matrix elements of the EMT to measurable parton distributions, we identify the non-local phase factor $\mathcal W$ with a straight Wilson line running along the LF direction $n$ to $r_n=r+\infty\, \eta n$, and then in the transverse direction to $\infty_\perp$. It is then clear that, beside the average target momentum $P=(p'+p)/2$ and the momentum transfer $\Delta=p'-p$, the matrix elements of the generic LF EMT also depend in principle on $N=\frac{M^2\,n}{P\cdot n}$ for a target of mass $M$, and on the direction of the Wilson line $\eta=\pm 1$. The scalar functions parametrizing the generic LF EMT are complex-valued functions of $\xi=-(\Delta\cdot N)/2(P\cdot N)$ and $t=\Delta^2$, which are the only two independent scalars that can be formed with $P$, $\Delta$ and $N$.

In Ref.~\cite{Lorce:2015lna}, we used the techniques presented in the Appendix A of Ref.~\cite{Meissner:2009ww} and found that the generic LF EMT for a spin-$1/2$ target can be parametrized as
$\langle p',S'|T^{\mu\nu}_a(0)|p,S\rangle =\overline u(p',S')\Gamma^{\mu\nu}_a(P,\Delta,N;\eta)u(p,S)$ with $a=1,\cdots, 5$ and the matrix $\Gamma^{\mu\nu}_a$ given by 
\begin{align}
\Gamma^{\mu\nu}_a&=Mg^{\mu\nu}A^a_1+\frac{P^\mu P^\nu}{M}\,A^a_2+\frac{\Delta^\mu\Delta^\nu}{M}\,A^a_3+\frac{P^\mu i\sigma^{\nu\Delta}}{2M}\,A^a_4+\frac{P^\nu i\sigma^{\mu\Delta}}{2M}\, A^a_5\nonumber\\
&+\frac{N^\mu N^\nu}{M}\,B^a_1+\frac{P^\mu N^\nu}{M}\,B^a_2+\frac{P^\nu N^\mu}{M}\,B^a_3+\frac{N^\mu i\sigma^{\nu\Delta}}{2M}\,B^a_4+\frac{N^\nu i\sigma^{\mu\Delta}}{2M}\, B^a_5+\frac{\Delta^\mu i\sigma^{\nu N}}{2M}\,B^a_6+\frac{\Delta^\nu i\sigma^{\mu N}}{2M}\, B^a_7\nonumber\\
&+\left[Mg^{\mu\nu}B^a_8+\frac{P^\mu P^\nu}{M}\,B^a_9+\frac{\Delta^\mu\Delta^\nu}{M}\,B^a_{10}+\frac{N^\mu N^\nu}{M}\,B^a_{11}+\frac{P^\mu N^\nu}{M}\,B^a_{12}+\frac{P^\nu N^\mu}{M}\,B^a_{13}\right]\frac{i\sigma^{N\Delta}}{2M^2}\nonumber\\
&+\frac{P^\mu\Delta^\nu}{M}\,B^a_{14}+\frac{P^\nu \Delta^\mu}{M}\,B^a_{15}+\frac{\Delta^\mu N^\nu}{M}\,B^a_{16}+\frac{\Delta^\nu N^\mu}{M}\,B^a_{17}+\frac{M}{2}\,i\sigma^{\mu\nu}\,B^a_{18}+\frac{\Delta^\nu i\sigma^{\mu\Delta}}{2M}\, B^a_{19}\nonumber\\
&+\frac{P^\mu i\sigma^{\nu N}}{2M}\,B^a_{20}+\frac{P^\nu i\sigma^{\mu N}}{2M}\, B^a_{21}+\frac{N^\mu i\sigma^{\nu N}}{2M}\,B^a_{22}+\frac{N^\nu i\sigma^{\mu N}}{2M}\, B^a_{23}\nonumber\\
&+\left[\frac{P^\mu\Delta^\nu}{M}\,B^a_{24}+\frac{P^\nu \Delta^\mu}{M}\,B^a_{25}+\frac{\Delta^\mu N^\nu}{M}\,B^a_{26}+\frac{\Delta^\nu N^\mu}{M}\,B^a_{27}\right]\frac{i\sigma^{N\Delta}}{2M^2}.\label{param}
\end{align}
For convenience, we introduced the notation $i\sigma^{\mu b}\equiv i\sigma^{\mu\alpha}b_\alpha$ and the factors of $i$ such that the real part of the scalar functions is $\eta$-even while the imaginary part is $\eta$-odd
\begin{equation}
X^a_j(\xi,t;\eta)=X^{e,a}_j(\xi,t)+i\eta\,X^{o,a}_j(\xi,t)
\end{equation}
as a consequence of time-reversal symmetry. The hermiticity property then implies that the real part of $B^a_j$ with $j\geq 14$ is $\xi$-odd and the imaginary part is $\xi$-even. For the other functions, it is the opposite. The fact that only 32 independent structures exist can be obtained from a naive simple counting : the EMT $T^{\mu\nu}_a$ has $4\times4=16$ components; the target state polarizations $\pm S$ and $\pm S'$ bring another factor of $2\times 2=4$; time-reversal and hermiticity having been used to fix the factors of $i$ and the $\xi$-dependence, we are left with parity which reduces the number of independent polarization configurations by a factor $2$. As announced, this leads to a total of $32$ independent complex-valued amplitudes.

Manifestly, the EMTs $T^{\mu\nu}_1$ and $T^{\mu\nu}_2$ do not depend on $N$ or $\eta$. All the corresponding scalar functions must then vanish except for $A^{e,a}_j(0,t)$ with $a=1,2$ and $j=1,\cdots,5$, which are linearly related to the standard energy-momentum form factors~\cite{Ji:1996ek,Bakker:2004ib,Leader:2013jra} as follows
\begin{equation}\label{EMFFs}
\begin{aligned}
A_q(t)&=A^{e,1}_2(0,t),&A_G(t)&=A^{e,2}_2(0,t),\\
B_q(t)&=A^{e,1}_4(0,t)+A^{e,1}_5(0,t)-A^{e,1}_2(0,t),&\qquad B_G(t)&=A^{e,2}_4(0,t)+A^{e,2}_5(0,t)-A^{e,2}_2(0,t),\\
C_q(t)&=A^{e,1}_3(0,t),&C_G(t)&=A^{e,2}_3(0,t),\\
\bar C_q(t)&=A^{e,1}_1(0,t)+\tfrac{t}{M^2}\,A^{e,1}_3(0,t),&\bar C_G(t)&=A^{e,2}_1(0,t)+\tfrac{t}{M^2}\,A^{e,2}_3(0,t),\\
D_q(t)&=A^{e,1}_4(0,t)-A^{e,1}_5(0,t),&0&=A^{e,2}_4(0,t)-A^{e,2}_5(0,t).
\end{aligned}
\end{equation}

\section{Linear and angular momentum constraints}\label{sec4}

The parametrization~\eqref{param} is only constrained by space-time symmetries. We briefly present here the additional constraints arising from the conservation of total linear and angular momentum.

The average four-momentum in the LF form of dynamics can be obtained by contracting the EMT with $\tfrac{1}{2M^2}\,N_\mu$ in the forward limit $\Delta\to 0$ 
\begin{equation}\label{momentum}
\langle p^\nu_a\rangle\equiv\frac{1}{2M^2}\,\langle P,S|T^{N\nu}_{a}(0)|P,S\rangle = P^\nu A^{e,a}_2+N^\nu (A^{e,a}_1+B^{e,a}_2)+\delta^{a3}\,\frac{\eta}{2}\,\epsilon^{\nu S}_T\,(B^{o,3}_{18}-B^{o,3}_{20}).
\end{equation}
Interestingly, the last term in Eq.~\eqref{momentum} originating from the potential EMT $T^{\mu\nu}_3$ is $\eta$-odd and can be interpreted as the spin-dependent contribution to the momentum arising from initial and/or final-state interactions, see \emph{e.g.}~\cite{Boer:2015vga} and references therein. It comes with the structure $\epsilon^{\nu S}_T\equiv\epsilon^{\nu\mu\alpha\beta}S_\mu n_\alpha \bar n_\beta$ where $\bar n$ is another lightlike four-vector satisfying $n\cdot\bar n=1$ and such that $P^\mu=(P\cdot n)\bar n^\mu+(P\cdot\bar n)n^\mu$, which means that a transverse target polarization is required. The total four-momentum being $P^\nu$, we recover from the sum over all partons the well-known momentum constraints
\begin{equation}\label{momcons}
\sum_{a=1,2}A^{e,a}_1(0,0)=\sum_{a=q,G}\bar C_a(0)=0,\qquad\sum_{a=1,2}A^{e,a}_2(0,0)=\sum_{a=q,G} A_a(0)=1.
\end{equation}

Thanks to the complete parametrization~\eqref{param}, we can easily compute the matrix elements of the corresponding OAM tensors $L^{\mu\nu\rho}_{a}=r^\nu T^{\mu\rho}_a-r^\rho T^{\mu\nu}_a$. Because of the explicit factors of $r$, the matrix elements of the generic LF OAM tensor need to be handled with care~\cite{Bakker:2004ib,Leader:2013jra}. For a longitudinally polarized target, we found a simple expression for the longitudinal component of OAM
\begin{equation}
\langle L^a_L\rangle\equiv\frac{\epsilon_{T\alpha\beta}}{2M^2}\left[i\,\frac{\partial}{\partial\Delta_\alpha}\langle p',S'|T^{N \beta}_{a}(0)|p,S\rangle\right]_{\Delta=0}= A^{e,a}_4(0,0).\label{OAM}
\end{equation}
Similarly, the quark and gluon spin contributions $S^{\mu\nu\rho}_1=\frac{1}{2}\,\epsilon^{\mu\nu\rho\sigma}\,\overline\psi\gamma_\beta\gamma_5\psi$ and $S^{\mu\nu\rho}_2=-2\uTr\!\left[G^{\mu[\nu}A^{\rho]}_\phys\right]$ can be expressed in terms of $L^{\mu\nu\rho}_4$ and $L^{\mu\nu\rho}_5$, respectively. We then found
\begin{equation}\label{spin}
\begin{aligned}
\langle S^q_L\rangle\equiv\frac{1}{2M^2}\,\langle P,S|\tfrac{1}{2}\,\epsilon_{T\alpha\beta}S^{N\alpha\beta}_1(0)|P,S\rangle&=-\frac{1}{2}\left[A^{e,1}_4(0,0)-A^{e,1}_5(0,0)\right],\\
\langle S^G_L\rangle\equiv\frac{1}{2M^2}\,\langle P,S|\tfrac{1}{2}\,\epsilon_{T\alpha\beta}S^{N\alpha\beta}_2(0)|P,S\rangle&=-A^{e,5}_4(0,0),
\end{aligned}
\end{equation}
where Eq.~\eqref{QCDEOM} has been used to express $L^{\mu\nu\rho}_4$ in terms of $T^{\mu\nu}_1$. Adding together the spin and OAM contributions, we naturally recover the Ji relation for total angular momentum~\cite{Ji:1996ek}
\begin{equation}\label{AMtot}
\begin{aligned}
\langle J^q_L\rangle&=\langle S^q_L\rangle+\langle L^q_L\rangle=\tfrac{1}{2}\left[A^{e,1}_4(0,0)+A^{e,1}_5(0,0)\right]=\tfrac{1}{2}\left[A_q(0)+B_q(0)\right],\\
\langle J^G_L\rangle&=\langle S^G_L\rangle+\langle L^G_L\rangle=\tfrac{1}{2}\left[A^{e,2}_4(0,0)+A^{e,2}_5(0,0)\right]=\tfrac{1}{2}\left[A_G(0)+B_G(0)\right].
\end{aligned}
\end{equation}
Finally, combining the fact that the total angular momentum is $1/2$ and the momentum constraints~\eqref{momcons}, we recover also the anomalous gravitomagnetic moment sum rule~\cite{Teryaev:1999su,Brodsky:2000ii}
\begin{equation}\label{cons3}
\sum_{a=1,2}[A^{e,a}_4(0,0)+A^{e,a}_5(0,0)-A^{e,a}_2(0,0)]=\sum_{a=q,G}B_a(0)=0.
\end{equation}

\section{Link with measurable parton distributions}\label{sec5}

The matrix elements of the EMT we are interested in can easily be expressed in terms of the GTMD correlator~\cite{Lorce:2012ce} 
\begin{equation}
\langle p',S'|T^{\mu\nu}(0)|p,S\rangle=\int\ud^4k\,k^\nu\,W^\mu_{S'S}.
\end{equation}
By considering appropriate projections~\cite{Meissner:2009ww,Lorce:2011dv,Lorce:2013pza}, we can at the end relate some of the scalar functions appearing in our generic parametrization~\eqref{param} to measurable parton distributions, like \emph{e.g.} Generalized Parton Distributions (GPDs) accessed in exclusive scatterings~\cite{Diehl:2003ny} and Transverse-Momentum dependent Distributions (TMDs) accessed in semi-inclusive scatterings~\cite{Collins:2011zzd}. The detailed relations between the EMT scalar functions and two-parton GPDs and TMDs of any twist can be found in~\cite{Lorce:2015lna}.

Among the interesting results, let us just mention that we derived the following sum rules
\begin{equation}\label{newSR}
\sum_{a=q,G}\int\ud x\,\ud^2k_T\,\tfrac{k^2_T}{2M^2}\,F^a(x,k^2_T)=0
\end{equation}
with $F^a=f^{\perp a}_{1T},f^{\perp a},f^{\perp a}_L,f^{\perp a}_{3T}$. They express the fact that the total transverse momentum (w.r.t. the target momentum) has to vanish. The leading-twist relation (\emph{i.e.} the one involving the Sivers function $f^{\perp a}_{1T}$) is known as the Burkardt sum rule~\cite{Burkardt:2003yg,Burkardt:2004ur}. The other three are new and much harder to test experimentally, but it would be very interesting to test them using phenomenological models, Lattice QCD and perturbative QCD.

\section{Conclusions}\label{sec6}

It is possible to define a gauge-invariant canonical energy-momentum tensor once one relaxes the assumption of strict locality. We argued that the canonical energy-momentum tensor can be considered as a physical object and \emph{a priori} measured experimentally \emph{via} particular moments of parton distributions extracted from various physical processes.

We presented here a complete parametrization for the matrix elements of the generic light-front energy-momentum tensor and discussed some of the constraints arising from linear and angular momentum conservation. We showed that this energy-momentum tensor can be related to particular moments of the generalized and transverse-momentum dependent parton distributions. Among the interesting results, we rederived in a simpler way the Burkardt sum rule and obtained three new sum rules involving higher-twist distributions, all expressing basically the conservation of transverse momentum. We expect in a near future exciting new developments in these matters coming from new experimental data obtained in existing and future facilities, and explicit investigations using phenomenological models, Lattice QCD and perturbative QCD.

\end{document}